\documentclass[12pt,preprint]{aastex}

\shorttitle{Transit Parallax}
\shortauthors{Scharf}

\begin{document}


\title{Exoplanet Transit Parallax}


\author{Caleb A. Scharf}
\affil{Columbia Astrobiology Center,
Columbia Astrophysics Laboratory, Columbia University, MC 5247, 550 West 120th Street, New York, NY10027, USA}
\email{caleb@astro.columbia.edu}



\begin{abstract}
The timing and duration of exoplanet transits has a dependency on observer position due to parallax. In the case of an Earth-bound observer with a 2 AU baseline the dependency is typically small and slightly beyond the limits of current timing precision capabilities. However, it can become an important systematic effect in high-precision repeated transit measurements for long period systems due to its relationship to secular perspective acceleration phenomena. In this short paper we evaluate the magnitude and characteristics of transit parallax in the case of exoplanets using simplified geometric examples. We also discuss further implications of the effect, including its possible exploitation to provide immediate confirmation of planetary transits and/or unique constraints on orbital parameters and orientations.

\end{abstract}


\keywords{planetary systems --- eclipses --- techniques:miscellaneous}


\section{Introduction}
Astronomical transits and parallax have a long history together. An excellent example is Edmund Halley's proposal that the solar transit of Venus could be used to estimate the parallax distance of the Sun by gathering data from observers at different locations on the Earth. Recently, the use of transit events to both detect and  characterize extrasolar planets (exoplanets)  has been wildly successful ~\citep{charbonneau00,castellano00,charbonneau02,charbonneau05}. In the next few years the transit detection of exoplanets will continue with ever-increasing sensitivity. Two space-based observatories (the {\sl Corot} mission of CNES and NASA's {\sl Kepler} mission) will be capable of detecting transiting Earth-sized planets approximately 1 AU from their parent stars, and a variety of ground-based efforts are ongoing.

Although transit measurements are currently corrected to the Heliocentric Julian Date (HJD) to account for finite light travel time differences with the Earth's orbital phase, they are not corrected for the (smaller) effect of parallax. While this should not have significantly biased current transit measurements, it may become increasingly important for both high-precision transit timing and the characterization of planets in orbits approaching 1 AU.  Parallax effects may also confuse efforts to utilize transit timing to characterize orbital precessions and those seeking the presence of other planetary objects in a system (e.g., ~\citet{agol05}) and moons of the transiting planet itself (e.g., ~\citet{doyle04}). However, parallax effects can also provide a new tool for characterizing exoplanet systems. In this paper we briefly state the nature and approximate magnitude of parallax effects and discuss their potential uses.

\section{Basic effects}

\subsection{Co-planar effects}

The simplest case to consider is that in which the orbital plane of the exoplanet system is co-planar with that of the Earth-Sun system. This situation is illustrated in Figure 1a. In this simplified schematic it is implicitly assumed that the orbital periods and phases of the Earth and exoplanet are such that the exoplanet will traverse the same piece of its orbit while the Earth is at either position 1 or 2. Obviously situations will arise where this is {\em not} true. For example, an exoplanet with an orbital period of precisely one year will always be observed from the same Earth position (for an Earth-based instrument), and no co-planar parallax effect will exist.

To first order (ignoring the finite stellar disk size and finite transit duration) the parallax angle $\theta$ is just given by $2R/d$, where $R$ is the orbital radius of the Earth and $d$ is the distance to the exoplanetary system. The transit timing (e.g., transit ingress or egress) difference $\Delta t$ between observers 1 and 2 (Figure 1a) is then simply given by:

\begin{equation}
\Delta t=\frac{R \; P}{\pi d}\;\;,
\end{equation}

where $P$ is the exoplanet orbital period. Thus, for example, the size of the effect for a transiting exoplanet with a $\sim 400$ day period in a system at 10 pc distance is $\Delta t\sim 5$ seconds. For a system at the same distance with period of precisely 0.5 years (i.e., one in which the transit will indeed occur as shown in Figure 1) then $\Delta t\sim 2.4$ seconds.

Relatively few transiting systems are expected within a distance of 10 pc owing to the geometric transit probability (e.g., ~\citet{sackett99}) and the low number of stars within this volume (between 300-400, mainly M-dwarfs). In order to monitor thousands to hundreds of thousands of stars, a typical search volume in our Galaxy extends from hundreds to a few thousand parsecs. In such cases, $\Delta t$ would only be a few $\sim 10^{-1} -10^{-2}$ seconds. 

The nominal (photon noise) 1-$\sigma$ timing error for transits of a 9th magnitude star over the mission lifetime of {\sl Corot} is approximately 7 seconds, for {\sl Kepler} it is approximately 1 second. For a 12th magnitude star, {\sl Corot}'s errors are in excess of a minute, while {\sl Kepler} attains approximately 6 seconds.

A rough estimate of the timing precision that can be achieved for transit measurements with a larger instrument can be made assuming photon counting statistics. \citet{holman05} point out that the uncertainty in the {\em central} transit time (assuming the ingress and egress are well sampled) $\sigma$, for a transit of duration $t$ is given by:

\begin{equation}
\frac{\sigma}{t} \sim (\Gamma t)^{-1/2} \left(\frac{R_p}{R_*}\right)^{-3/2}\;\;\;,
\end{equation}

where $R_p$ and $R_*$ correspond to the planet and star radius respectively. $\Gamma$ is the actual photon count rate of the star in the given instrument. While this assumes the simplest direct approach to estimating the central transit time it nonetheless provides an indication of the inherent noise level. For a typical 8-m class telescope $\Gamma$ is estimated to be $\sim 5.5 \times 10^{10} \cdot 10^{-0.4(V-12)}$ hour$^{-1}$. Thus, for transit durations of the order of a few hours and a stellar magnitude $V=12$, then measurements of a Jupiter sized planet (assuming a solar mass parent star) would  yield $\sigma \sim 0.3$ seconds and an Earth-sized planet $\sigma \sim 10$ seconds. 

Thus, while $\Delta t$'s of the level described are beyond the sensitivity of {\sl Corot} and {\sl Kepler}, for 8-m class instruments and Jupiter sized planets around 12th magnitude or brighter stars they begin to enter the realm of practical measurement. The above estimates are also based upon straightforward estimation of the transit curve timing. Observational strategies which exploit an iterative approach (i.e. making use of the prior knowledge of previous transit events) to determine the optimal start and end of a given exposure may improve on this somewhat. However, we also note that other systematics, including orbital precession, stellar variation, and instrumental variations (e.g. thermal conditions) will need to be very well understood to reach the level of
$\sim 0.1$ second timing precision.

\subsection{Other geometries}

If the exoplanet-star plane is normal to that of the solar system (Figure 1b) then parallax alters the transit duration by altering the effective inclination, or latitude ($\delta$) of the transit . The general form of the transit duration ($\tau$) for a circular orbit is given by ~\citep{sackett99}:

\begin{equation}
\tau=\frac{P}{\pi} \sin^{-1}\left[\frac{R_*}{a}\left[\frac{(1+\frac{R_p}{R_*})^2 -(\frac{a \cos i}{R_*})^2}{1-\cos^2 i}\right]^{1/2}\right]\;\;,
\end{equation}

where $R_*$ and $R_p$ correspond to the parent star and exoplanet radii respectively, $a$ is the orbital radius of the planet, and $i$ is the orbital inclination of the exoplanet. In Figure 2 the transit duration for a Jupiter-sized planet orbiting a 1 $M_{\odot}$, 1 $R_{\odot}$ star is plotted versus inclination angle for a variety of orbital periods/radii. As is well known, longer period planet transits will only be seen at inclinations close to 90$^{\circ}$, but the transit duration may be quite long. In Figure 3 we plot the {\em change} in transit duration ($\Delta \tau$) between observers separated by 2 AU (e.g., Figure 1b),  as a function of the transit duration (which itself is a function of inclination as in Figure 2). Several cases for different orbital periods (radii) and system distance are shown. The $\Delta \tau$ effect is, of course, largest for inclinations at the steepest parts of the curves in Figure 2. At distances of 100 pc (which are arguably the most realistic for the majority of
transit studies) $\Delta \tau/\tau$ ranges between approximately $0.001$\% (10 day period, corresponding to sub 0.1 second duration variation) and (rarely) as much as 10\% (1000 day period, corresponding to $\sim 1 $ minute duration variation). For periods of 300 days the range is 0.003 to 5\% at 100 pc, corresponding to time variations of $\sim 0.5$ seconds to $\sim 10$ seconds.

Thus, in this geometry the potential variations in transit duration are almost within the grasp of instruments such as {\sl Corot} or {\sl Kepler}, and certainly 8-m class observatories (assuming systematics can be overcome).
Variations of this order can systematically bias estimates of planet size (e.g. \citet{minniti07}).

Between this perpendicular configuration and the co-planar case are intermediate orientations of the Earth-Sun and exoplanet-star planes. In these cases the transit timing and duration variation can, in principle, provide constraints on  the relative orientation (not just inclination) of an exo-planet system and the solar system.  We also note that the signatures of ring structures in transit curves (e.g., \citet{barnes04}) can also exhibit a parallax effect, owing to the  dependency on both ring inclination and the system inclination.

\section{Relative motions}

There are other potential sources of systematic  error in high-precision transit timing experiments which become increasingly important as longer period systems are studied and as systems are monitored over extended times (i.e. multiple orbital periods). In the local stellar neighborhood, relative stellar velocities are of the order $\sim 20$ km s$^{-1}$ (e.g., ~\citet{bienayme99}) corresponding to an annual relative motion of about 4 AU. In the case of pure line-of-sight relative motion the stellar redshift can be used to predict the sense and magnitude of this light-travel time effect - and for long period transit observations the apparent transit time will therefore shift in an appreciable, but determined way. In the case of tangential relative motion, unless very high-precision astrometry is available, an unknown systematic drift will be present that will act in the same way as the transit parallax described above, except as a secular phenomenon akin to secular perspective acceleration (e.g. \citet{kamp67,dravins99}). Indeed, if such a systematic drift in timing and/or transit duration over several years could be measured, it could be used as a direct surrogate for an astrometric proper motion. 

In the co-planar geometric case (Equation 1) a 400 day period planet at a distance of 100 pc with a relative motion to the observer of 4 AU  per year (see above) would exhibit a secular variation in transit timing of some 10 seconds over a period of ten years. In the case of an apparent inclination drift (Figure 3), a similar system at 100 pc observed over a ten year period could exhibit a transit duration change of as much as $\sim 20$ seconds to 3 minutes (depending on the initial inclination). These are effects that therefore need to be included in any long term monitoring programs of transits, especially those aimed at detecting other secular variations, such as orbital precession, and even the detection of unseen planetary, or satellite, companions. In the case of planetary companions, variations between seconds and minutes are expected - with a variety of characteristics (e.g. \citet{holman05,agol05}). For satellite companions, systems such as that represented by Jupiter-Europa involve 6 second timing variations, while the Earth-Moon system exhibits variations of $\sim 2.5$ minutes (e.g. \citet{sartoretti99,doyle04}).

Alternatively, if high precision astrometric data is available (e.g. from future instruments such as {\sl SIM} or {\sl GAIA}, see below) and the three-dimensional relative motion is well constrained, then the observed transit duration and timing drift can be modeled so as to obtain constraints on the full geometric orientation of the system with respect to the Earth-Sun system.

\section{Exploiting parallax}

Upcoming and future planet transit experiments involve single observatories or instruments. There is however an advantage to simultaneous observation of transit events with widely separated instruments, in particular, space-based platforms with very large baselines. We discuss here, briefly, some of the additional science that would be enabled by such a hypothetical experiment.

As much as the transit parallax is a potential bias for Earth-based observations, it may also provide an opportunity for obtaining new constraints
on exoplanet systems. Here we briefly consider a scenario in which a pair of spacecraft in a solar orbit of at least 1AU radius, but separated by 180$^{\circ}$, are used to simultaneously monitor star-planet transits (with timing corrected to HJD).

As shown above, parallax creates a relatively unambiguous signature during a transit. For a star monitored by both spacecraft instruments in the case illustrated by Figure 1a, an exoplanet transit will
exhibit a distinct timing variation for both ingress and egress. This can be used to uniquely identify just {\em one} transit event as being due to an exoplanet, and will clearly differentiate it from either stellar microvariability or eclipsing binaries. 

If the physical distance to the parent star is known, then the timing variation can also be used to estimate the orbital velocity during transit. In the future {\sl GAIA} will produce $\sim 25\; \mu$arcsec precision parallax measurements for some hundreds of millions of stars, and {\sl SIM} will produce $1-4\;\mu$arcsec precision parallaxes for tens of thousands of stars.
With such distance measurements in hand then timing variation of a {\em single transit} will provide an estimate (assuming a circular orbit) of the orbital radius or period (if the stellar mass is known). Once further transits are seen, and the period is properly known, this information can be further used to constrain the orbital eccentricity by comparing the
sampled velocity to that expected for a circular orbit. While the longitude of periastron is not known, this can nonetheless place a broad limit on eccentricity. In the most extreme cases, where the transit is seen during the exoplanet's periastron or apastron the ratio of the equivalent period circular orbital velocity to the actual velocity is given by $[(1-e)/(1+e)]^{1/2}$ and $[(1+e)/(1-e)]^{1/2}$ respectively. As an example, for an orbital eccentricity of $e=0.3$ this translates to a velocity change of some $+27$\% and $-36$\% respectively, relative to the circular orbital velocity for the same orbital period. In fact, both the ingress timing and egress timing difference seen by the two spacecraft could be exploited in this fashion - although with less sensitivity.

Thus, armed with a high-precision measurement of the orbital period, the simultaneous long baseline observation of a co-planar transit will both determine the sign of the deviation from circularity (and hence determine whether periastron or apastron is closest to the observer) and place a {\em lower limit} on the orbital eccentricity (e.g., using the more general form of transit duration given by ~\citet{tingley05}). 
Such constraints may be very hard, if not impossible, to make for terrestrial mass planets at $\sim 1$ AU using other techniques. The apparent intrinsic limits of radial velocity measurements at $\sim 1$m s$^{-1}$  preclude the use of this method.

Finally we note that although vastly more technically challenging, if the {\em secondary eclipse} could be detected by the same instruments (e.g., for giant planets in the IR), a similar constraint could be made for the antipodal orbital velocity which, together with the primary eclipse data, and the relative timing of both primary and secondary eclipses, would enable the orbital parameters to be fully determined. For low-mass planets at $\sim 1AU$ this is something that might otherwise remain beyond the capabilities of Doppler-based measurements.

Obviously, if transit timing perturbations due to the presence of unseen
planets or objects are present they can confuse this scheme. However, as described previously, the parallax effects can in principle be well understood and this simultanenous observation will separate the phenomena.

\subsection{Constraining stellar limb models}

In addition to providing strong confirmation of planet transits and constraints on orbital parameters, the use of simultaneous long-baseline observations might in principle also provide information on the transit light curve shape. In the case illustrated in Figure 1b the transit occurs at different inclinations for two observers and hence at different stellar latitudes. The stellar limb darkening is the dominant factor in determining the shape of the transit light curve. Realistically however the physical range of the stellar photosphere sampled with long-baseline observations is likely to be small. At 10 pc with a 2 AU total baseline the actual difference in sampled photosphere for a solar radius star is only some $\sim 0.7$ km.
Thus, although there is an effect, it is very unlikely to be measurable.

\section{Summary and Discussion}

The effect of parallax on exoplanet transit observations is both a potential source of systematic error in quantifying transit events, and a possible means by which more information can be extracted from a transit. Although the magnitude of the effect for an Earth-bound observer is, in most forseeable cases, quite small it may nonetheless enters the regime of future high precision transit measurements. In particular, the secular
changes in transit timing and duration due to relative stellar motion should be taken into account for high-precision, long timeline transit observations. This amounts to an additional systematic that might confuse efforts to measure orbital precessions and the presence of other planets in a system. An observational scheme with simultaneous measurements of transits made across a long baseline (e.g., 2 AU) could provide unique constraints on the orbital parameters of Earth-type transiting planets. 

Extremely high photometric precision would be needed in order to perform the required timing to detect a 2 AU baseline transit parallax effect. However, together with the utility of precision transit timing experiments in detecting the presence of other planets or moons in a system (\citet{sartoretti99,holman05,agol05,doyle04}), the potential of the transit parallax phenomenon suggests that efforts to achieve new levels of photometric timing precision could be highly productive.

\acknowledgments The author acknowledges the funding support of the Columbia Astrobiology Center through
Columbia University's Initiatives in Science and Engineering, and the
support of the Columbia Astrophysics Laboratory. This work is also
directly supported by a NASA Astrobiology: Exobiology and Evolutionary Biology; and Planetary Protection Research grant, \# NNG05GO79G. David Spiegel and David Helfand are thanked for useful discussions. The referee is thanked for comments which helped improve this manuscript.

\clearpage

\begin{figure}
\plotone{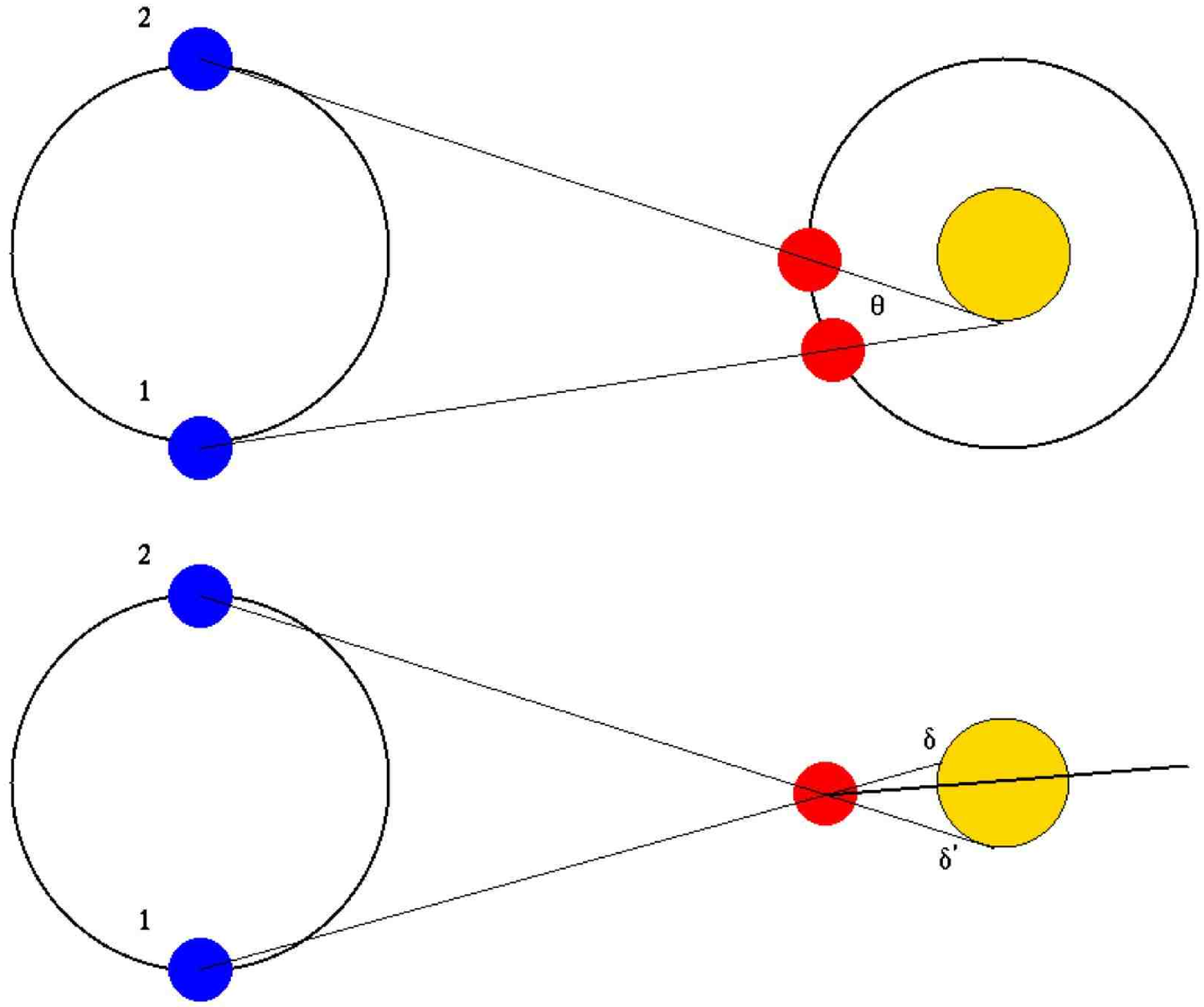}
\caption{The two fundamental geometries relevant to exoplanet transit parallax are illustrated schematically. {\it Upper plot}: Co-planar alignment of the Earth-Sun system (left, Sun not shown) and exoplanet system (right, exoplanet and parent star shown). Clockwise orbital motion is assumed.  A transit observed when the Earth is at location 1 will occur earlier than when observed from location 2. {\it Lower plot}: Perpendicular alignment between the Earth-Sun system and the exoplanet system. The exoplanet orbit is shown as a straight line with a small inclination. Depending on the location of the observer (position 1 or 2) the apparent inclination, or latitude $\delta$, of the transit will vary.}
\end{figure}

\begin{figure}
\plotone{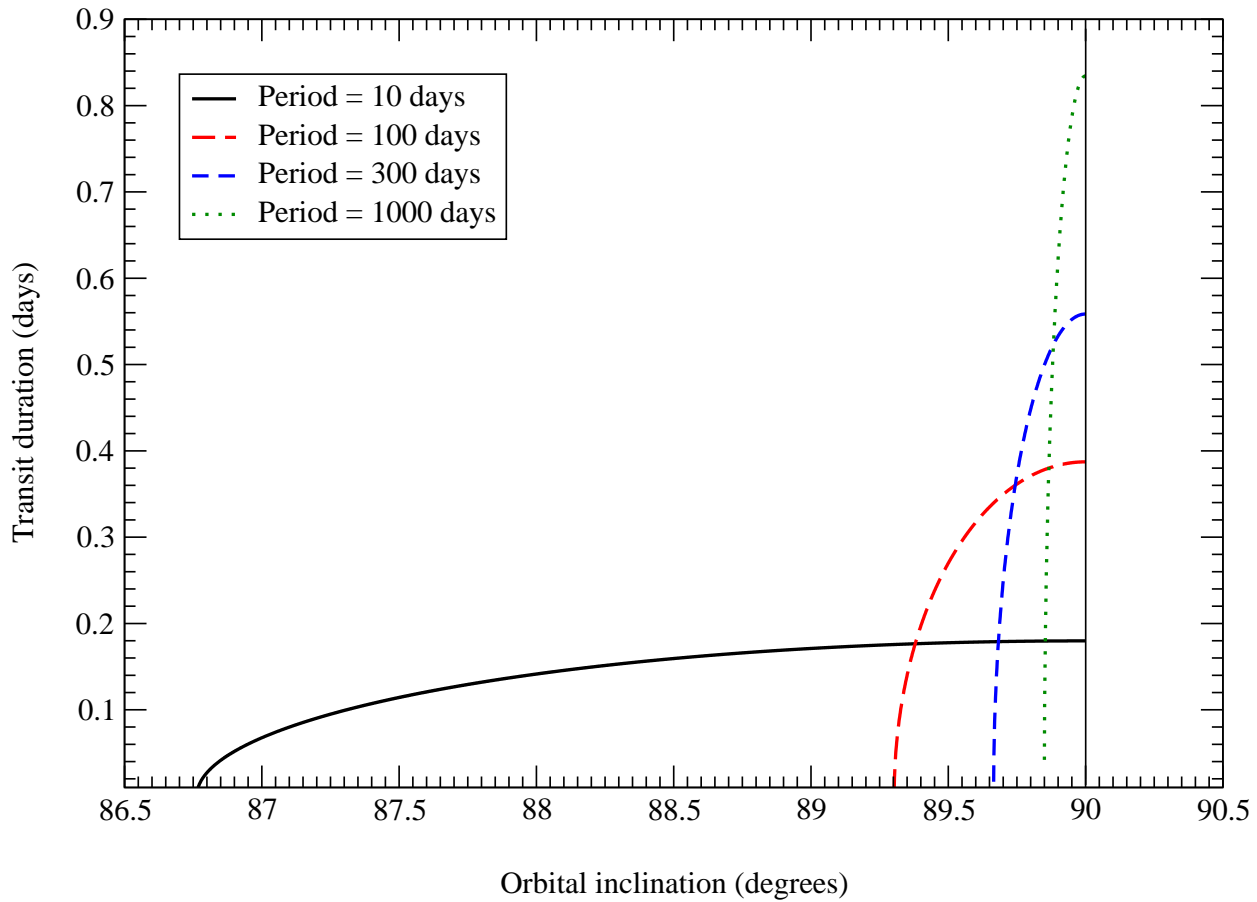}
\caption{The transit duration of a Jupiter radius planet orbiting a 1 $M_{\odot}$ , 1 $R_{\odot}$ star is plotted as a function of orbital inclination for a range of orbital periods (Equation 2). The range of inclinations over which transit detections can be feasibly made is a strongly decreasing function of period (increasing orbital radius). At inclinations approaching critical ``cut-off" values, the steepness of this function results in a high sensitivity of duration to inclination, and a maximal parallax effect (Figure 3). }
\end{figure}

\begin{figure}
\plotone{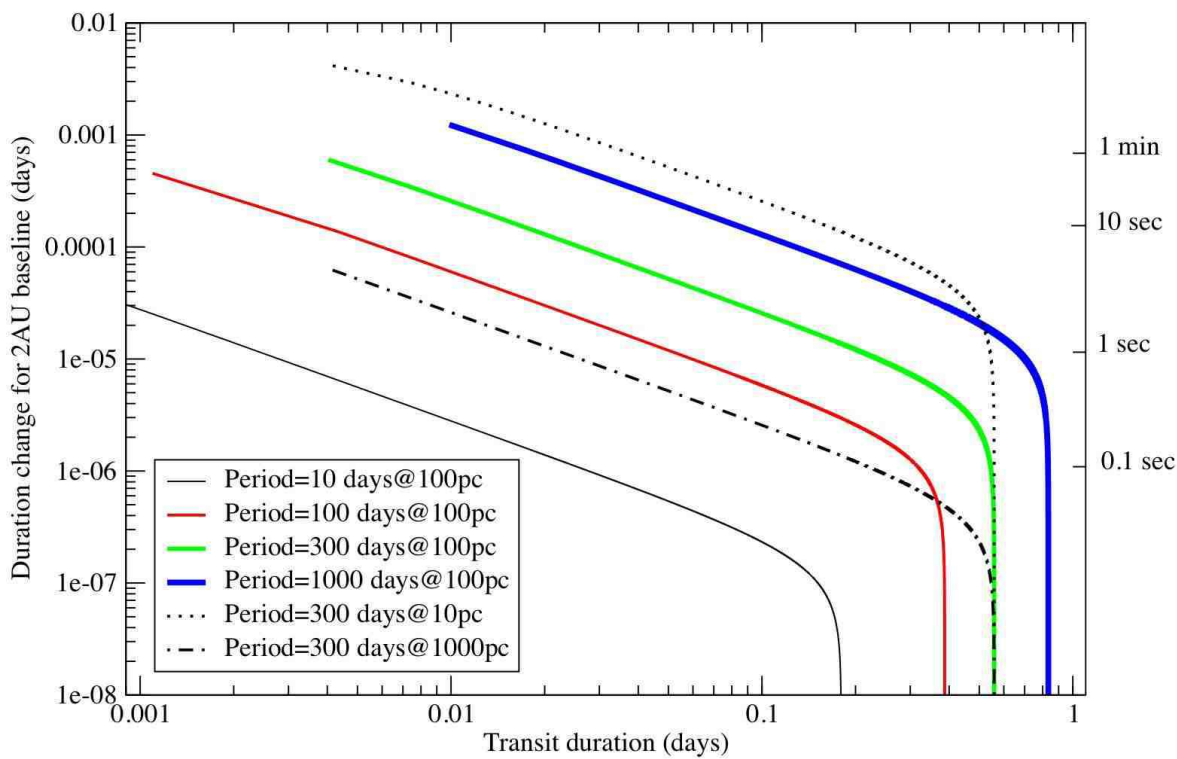}
\caption{The change in apparent transit duration between observers on a 2 AU baseline is plotted versus total transit duration for a Jupiter radius planet at a range of orbital periods (assuming a 1 $M_{\odot}$, 1 $R_{\odot}$ parent star). 
Inclination varies along each curve from $90^{\circ}$ at the right-hand ``knee" in each curve, and decreases toward the left according to Figure 2. The dotted curve assumes a system at a distance of 10 pc, the dot-dashed curve assumes a system 1000 pc distant; all other curves assume a 100 pc distance.}
\end{figure}

\end{document}